\documentclass[12pt]{iopart}

\usepackage{iopams}  
\usepackage[breaklinks=true,colorlinks=true,linkcolor=blue,urlcolor=blue,citecolor=blue]{hyperref}
\usepackage{geometry}
\usepackage{graphicx}
\usepackage{caption}
\usepackage{subcaption}
\usepackage{cite}
\usepackage{tabularx}
\usepackage{color}
\usepackage{xfrac}
\begin{document}

\title{Spatial patterns and biodiversity in rock-paper-scissors models with regional unevenness}

\author{J. Menezes\textsuperscript{1,2} and M. Tenorio\textsuperscript{2}}

\address{$^1$ Institute for Biodiversity and Ecosystem
Dynamics, University of Amsterdam, Science Park 904, 1098 XH
Amsterdam, The Netherlands}

\address{$^2$ School of Science and Technology, Federal University of Rio Grande do Norte\\
59072-970, P.O. Box 1524, Natal, RN, Brazil}

\ead{jmenezes@ect.ufrn.br}



\begin{abstract}
Climate changes may affect ecosystems destabilising relationships among species. We investigate the spatial rock-paper-scissors models with a regional unevenness that reduces the selection capacity of organisms of one species. Our results show that the regionally weak species predominates in the local ecosystem, while spiral patterns appear far from the region, where individuals of every species play the rock-paper-scissors game with the same strength. Because the weak species controls all local territory, it is attractive for the other species to enter the local ecosystem to conquer the territory. However, our stochastic simulations show that the transitory waves formed when organisms of the strong species reach the region are quickly destroyed because of local strength unbalance in the selection game rules. Computing the effect of the topology on population dynamics, we find that the prevalence of the weak species becomes more significant if the transition of the selection capacity to the area of uneven rock-paper-scissors rules is smooth. Finally, our findings show that the biodiversity loss due to the arising of regional unevenness is minimised if the transition to the region where the cyclic game is unbalanced is abrupt. Our results may be helpful to biologists in comprehending the consequences of changes in the environmental conditions on species coexistence and spatial patterns in complex systems.
\end{abstract}

\section{Introduction}

Spatial interactions among species are responsible for the rich biodiversity found in nature \cite{ecology,Nature-bio}. Natural selection leads to the dominance 
of the species that adapt the most to the local conditions, creating a variety of ecosystems \cite{darwin,island1,bioislands}. 
There is plenty of evidence that the environment plays a vital role in population dynamics, affecting organisms' mobility and predation ability. It has been reported that spatial variability of competition conditions impacts coexistence, defining species richness at local and regional scales \cite{CHESSON1985263,doi:10.1086/323589,https://doi.org/10.1890/02-0528}. The geographic dependence of species interactions has been observed, for example, in the latitudinal variation of ecological richness, which may induce diet adaptation in the Artic \cite{https://doi.org/10.1002/ece3.1980}. Environmental conditions may also unbalance stable ecological relationships, resulting in a regional species predominance that does not occur in other geographic regions \cite{10.1093/aobpla/plw081,geographic,10.3389/fmars.2020.567758,plantar,urbanisation,size}. Because of this, climate changes can locally affect the relationship among species resulting in biodiversity loss due to the unbalance of the strength they compete with each other for natural resources \cite{climate,climate2}.

Experiments with bacteria \textit{Escherichia coli} revealed a cyclic nonhierarchical dominance among three strains, whose spatial interactions can be described by rules of the rock-paper-scissors model \cite{Coli,bacteria,Allelopathy}. 
In this famous game,  scissors cut paper, paper wraps rock, rock crushes scissors, describing the selection interaction among species \cite {Directional1, Directional2}. The same cyclic competition has been observed in systems of lizards and coral reefs \cite{lizards,Extra1}. 
Because of the relevance of the rock-paper-scissors model to describe biological systems, we study a three-species model where individuals are impacted by environmental conditions that determine their capacity to compete in the cyclic game. 

We consider the existence of a region where one out of the species is strongly affected, 
inducing an unevenness in the spatial rock-paper-scissors selection rules. The investigation is performed by running sthochastics simulations, widely used in literature to study biological systems \cite{Reichenbach-N-448-1046,Szolnoki-JRSI-11-0735, Moura, Anti1,anti2,MENEZES2022101606,PhysRevE.97.032415,Avelino-PRE-86-036112,pairwise1,Nagatani2018,ham,PhysRevE.99.052310,TENORIO2022112430,tanimoto2}. 
We explore the geometric parameters of the area where one species is weakened and quantify the influence of the transition between spatial interactions within the region and far from it, where all organisms compete with the same strength. 

As reported by some authors, the spatial structure may influence population dynamics, determining the chances of species to coexist \cite{park22,SCHREIBER20131,neigh,directionp}. Furthermore, the unbalancing selection, reproduction, and mobility interactions may provoke species extinction if the unbalacing in the spatial interactions is significantly strong \cite{weakest,uneven,PedroWeak,Weak4,AVELINO2022111738}. Although the relevance of studies of cyclic game systems of out equilibrium,
the dynamics of spatial patterns and the effects on biodiversity have not been in the case of the unevenness in the rock-paper-scissors rules are out equilibrium not 
exclusively in part of space. Instead, it has been considered that all organisms of one species are weak because of an external cause, for example, a disease that reaches the systems undermining the selection capacity of all organisms of one species. In our model, the weakening is spatially structured, with an organism only becoming debilitated when approaching one specific region.

Our goal is to describe the pattern formation process, quantifying the dynamics of species densities within and outside the selection-limiting region. 
Since coexistence may be jeopardised if the disequilibrium in the spatial interactions is significant, we also aim to quantify the impact of the regional unevenness in biodiversity, calculating the coexistence probability due to the relevance of this issue to biodiversity conservation \cite{uneven,Park2017,PARKCHAOS,PhysRevE.105.014215,MENEZES2022101028,10.1063/5.0106165,MENEZES2023113290}.

The outline of this paper is as follows: the model and our methods are introduced in 
Sec.~\ref{sec2}, where we present our simulations and define the parameters. In Sec.~\ref{sec3}, we study the organisms' spatial organisation, highlighting the details of the pattern formation.
We the effects of the regional unevenness in the species populations in Sec.~\ref{sec4}. 
The coexistence probability in terms of the model parameters are presented in Sec.~\ref{sec5}.
Finally, we present our conclusions in Sec.~\ref{sec6}.

\section{Model and methods}
\label{sec2}

\begin{figure}[t]
\centering
\includegraphics[width=40mm]{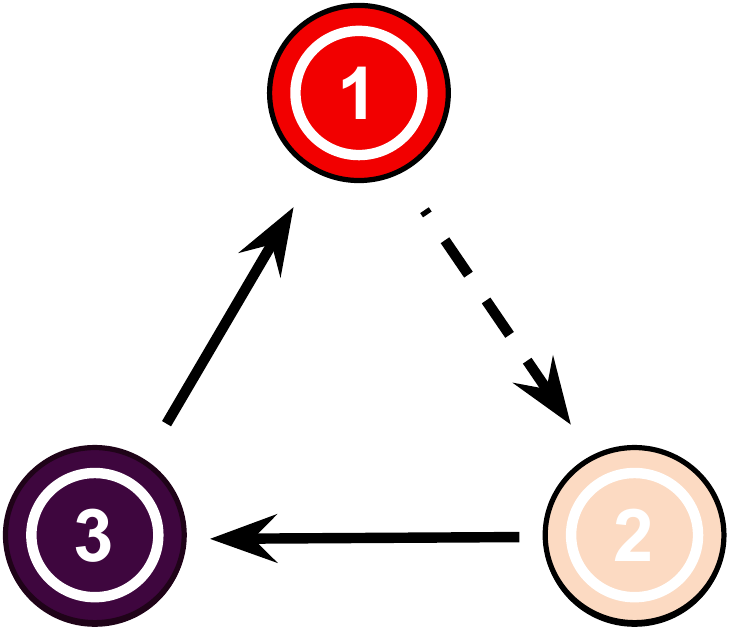}
\caption{Illustration of selection rules in the rock-paper-scissors model. The arrows describe the cyclic dominance of individuals of species $i$ over organisms of species $i+1$. The dashed arrow indicates that only individuals of species $1$ become regionally weak due to the vulnerability to environmental effects.
}
\label{fig1}
\end{figure}

\begin{figure}[h]
 \centering
       \begin{subfigure}{.4\textwidth}
        \centering
        \includegraphics[width=45mm]{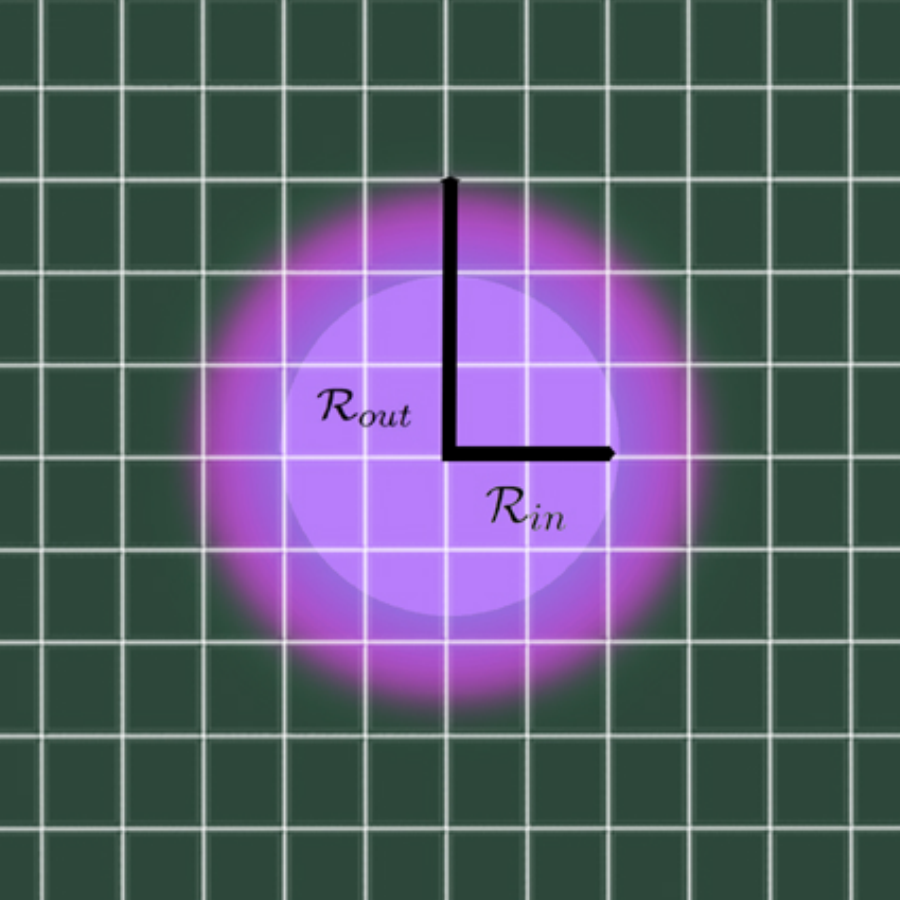}
        \caption{}\label{fig2a}
    \end{subfigure}
   \begin{subfigure}{.4\textwidth}
        \centering
        \includegraphics[width=75mm]{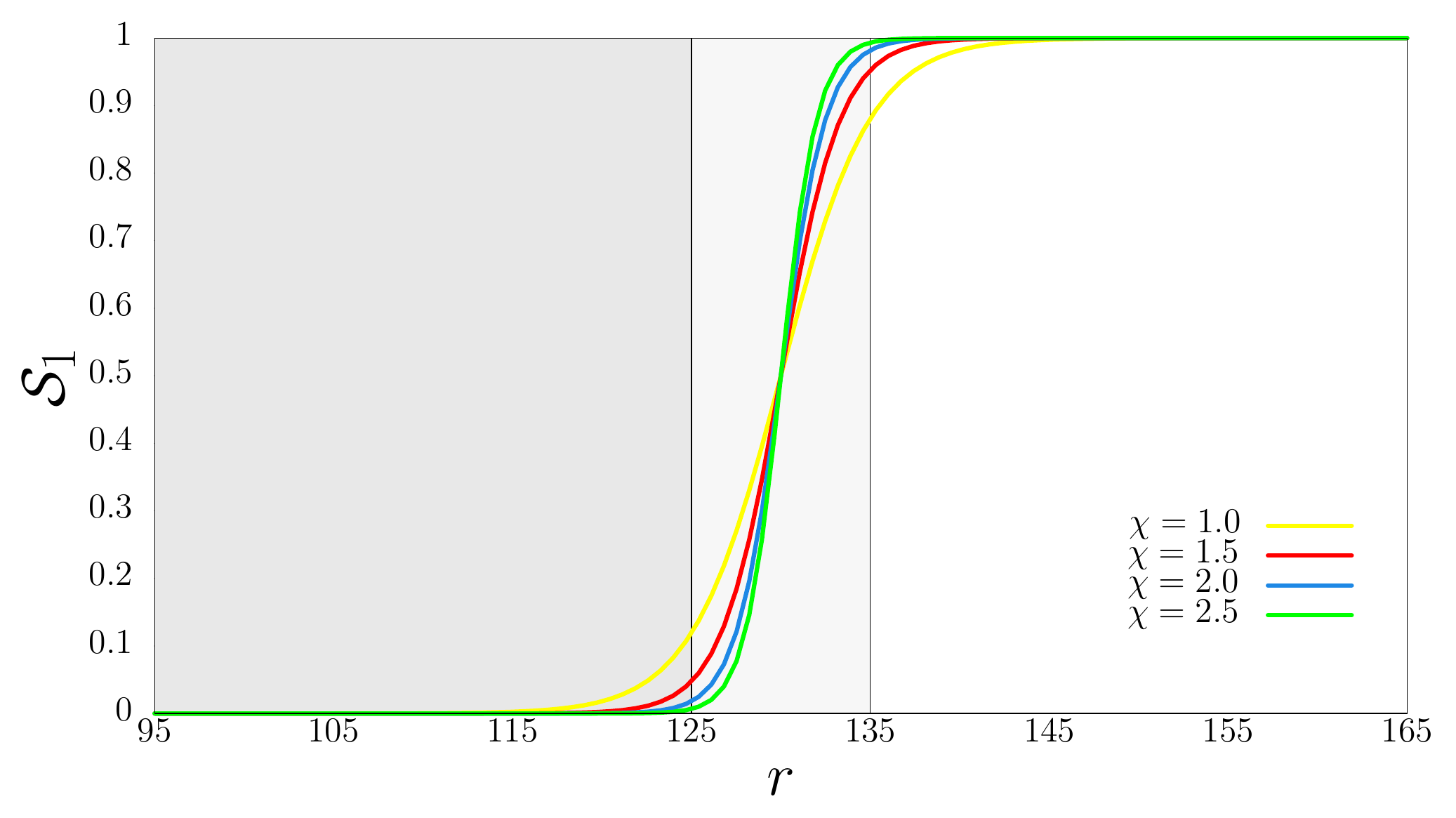}
        \caption{}\label{fig2b}
    \end{subfigure} 
\caption{Implementation of the region of the uneven cyclic game.
In Fig.~\ref{fig2a}, the light purple disk of radius $\mathcal{R}_{in}$ represents the region where organisms of species $1$ compete with minimum selection capacity, while outside the circle with radius $\mathcal{R}_{out}$, the selection capacity is maximum (green area). The circular strip of width $\mathcal{R}_{out}-\mathcal{R}_{in}$ shows the transition area where species $1$ is partially affected. Figure \ref{fig2b} shows the crossover between the minimum and maximum selection capacity of species $1$ according to Eq.~\ref{eq1}, for $\mathcal{R}_{in}=125$, $\mathcal{R}_{out}=135$, computed for various values of the transition parameter $\chi$. For all individuals of species $2$ and $3$, the selection capacity is maximum, independently of the their spatial position: $\mathcal{S}_2 = \mathcal{S}_3= \mathcal{S}_1^{max}$.}
  \label{fig2}
\end{figure}

We study a spatial version of the rock-paper-scissors game, where organisms of one out of the species are affected by environmental conditions when inside a geographic area, thus, totally or partially losing selection capacity. Our model is composed of three species, which we identify with the notation $i$ with $i= 1,...,3$, and the cyclic identification $i=i\,+\,3\,\alpha$ where $\alpha$ is an integer. The selection dominance are shown by the arrows in the illustration in Fig.~\ref{fig1}, where individuals of 
species $i$ eliminate individuals of species $i+1$. We consider that only organisms of species $1$ regionally face difficulties in selecting organisms of species $2$, as illustrated by the dashed arrow in Fig.~\ref{fig1}.

We perform stochastic simulations to describe the space-time dynamics of three species whose individuals interact with immediate neighbours. In contrast with the Lotka-Volterra model, where each grid site is always occupied by an individual of one of the species (conservation law for the total number of particles), in the May-Leonard model, selection interactions produce empty spaces \cite{Lotka,Volterra,leonard}. This leads to the emergence of spiral patterns whose characteristic length scales are associated with the parameters \cite{LV-ML}.
Our numerical implementation follows the May-Leonard framework, widely assumed to investigate spatial games \cite{Reichenbach-N-448-1046,Avelino-PRE-86-036112,driven}. We build
square lattices with periodic boundary conditions, where each grid point contains at most one individual, which means that the maximum number of organisms is $\mathcal{N}$, the total number of grid points. 

First, we allocate one individual at a random grid point to build the initial conditions.
Then, the interactions start being stochastically implemented as follows:
\begin{itemize}
\item 
Selection: $ i\ j \to i\ \otimes\,$, with $ j = i+1$, where $\otimes$ means an empty space: a selection interaction eliminates the organism of species $i+1$, generating an empty space; 
\item
Reproduction: $ i\ \otimes \to i\ i\,$: every time one reproduction is realised, a new organism of species $i$ occupies the available empty space.
\item 
Mobility: $ i\ \odot \to \odot\ i\,$, where $\odot$ means either an organism of any species: an organism of species $i$ exchanges positions with either another individual of any species or an empty space.
\end{itemize}
Selection, reproduction and mobility interactions are chosen with probabilities $s$, $r$ and $m$, respectively, which are the same for all individuals of every species. The interactions are implemented following the Moore neighbourhood, i.e., individuals may interact with one of their eight nearest neighbours.

Our algorithm follows the steps: i) an active individual is randomly chosen among all organisms of every species; ii) one of the possible interactions is raffled to be executed; 
iii) one of the eight immediate neighbour positions is randomly raffled to suffer the sorted interaction. The execution is always well-succeeded in the case of mobility interaction; however, selection and reproduction are constrained to the content of the passive position. If the interaction is implemented, one timestep is counted. Otherwise, the steps are redone. Our time unit is called generation, the necessary time to $\mathcal{N}$ timesteps to occur.


We introduce the selection capacity, $\mathcal{S}_i$, with $i=1,2,3$, as the probability of an individual of species $i$ killing an organism of species $i+1$, when the selection interaction is stochastically chosen. In our model, $\mathcal{S}_i$ is a function of the space, meaning that an individual's spatial position determines the success of selection interaction. 
Throughout this paper, we consider that only individuals of species $1$ are affected by regional selection restrictions, with $\mathcal{S}_1$ varying according to the individual's position. 

The regional selection unevenness is implemented by defining
two concentric circular areas to implement a regional reduction in the organisms' selection capacity, as illustrated in Fig.~\ref{fig2a}. 
Accordingly, 
\begin{itemize}
\item
Light purple region: inside the disk of radius $\mathcal{R}_{in}$, the selection capacity of individuals of species $1$ is minimum.
\item
Dark green area: outside the disk of radius $\mathcal{R}_{out}$, $\mathcal{S}_1$ is maximum.
\item
Gradient purple circular strip: in the transition area, the selection capacity is a crossover between the minimum and the maximum value.
\end{itemize}

For an individual of species $1$ located at a distance $\mathcal{R}$ from the centre of circular region, the selection capacity is given by
\begin{equation}
\mathcal{S}_1 =\frac12 \, \left\{1+\tanh\left[\chi\left(\frac{2\,\mathcal{R}-\mathcal{R}_{out}-\mathcal{R}_{in}}{\mathcal{R}_{out}-\mathcal{R}_{in}} \right)\right]\right\},
\label{eq1}
\end{equation}
where $\chi$ is the transition parameter, with $\chi \geq 0$, representing the crossover between $\mathcal{S}_1$ within the selection-limiting area, $\mathcal{S}_1 =(1-\tanh \chi)/2$,
and in the area without restriction in the selection activity, $\mathcal{S}_1 =(1+\tanh \chi)/2$. Figure \ref{fig2b} shows the crossover between the minimum and maximum $\mathcal{S}_1$ for $\chi=1.0$ (yellow line), $\chi=1.5$ (red), $\chi=2.0$ (blue line), and $\chi=1.0$ (green line). The dark grey area represents the region within the disk of radius $\mathcal{R}_{in}$. At the same time, the white panel shows the region without selection limitation for species $1$ ($r>\mathcal{R}_{out}$); the light grey interval shows the distance to the disk centre where an organism of species $1$ partially suffer the reduction in selection capacity. 
For all organisms of species $2$ and $3$, the selection capacity is maximum, independently of the spatial position; this means that  
$\mathcal{S}_2 = \mathcal{S}_3 = \mathcal{S}_1^{max} = (1+\tanh \chi)/2$,
which is the same as $\mathcal{S}_1$ far from the disk of radius $\mathcal{R}_{out}$.

The population dynamics is quantified by calculating the spatial densities $\rho_i$, i.e., the fraction of the grid occupied by individuals of the species $i$, that is a function of time $t$, i.e., $\rho_{i}(t) = I_i(t)/\mathcal{N}$, where $I_i(t)$ is the total number of individuals of species $i$ at time $t$.
Initially, every species occupies the same fraction of the grid, with $I_i(t) \approx \mathcal{N}/3$, for $i=1,2,3$.


\begin{figure*}
\centering
    \begin{subfigure}{.24\textwidth}
        \centering
        \includegraphics[width=33mm]{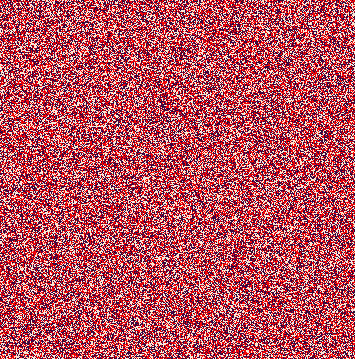}
        \caption{}\label{fig3a}
    \end{subfigure} %
       \begin{subfigure}{.24\textwidth}
        \centering
        \includegraphics[width=33mm]{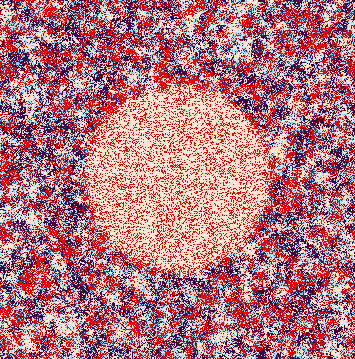}
        \caption{}\label{fig3b}
    \end{subfigure} %
   \begin{subfigure}{.24\textwidth}
        \centering
        \includegraphics[width=33mm]{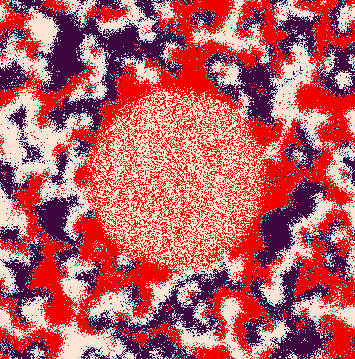}
        \caption{}\label{fig3c}
    \end{subfigure} 
   \begin{subfigure}{.24\textwidth}
        \centering
        \includegraphics[width=33mm]{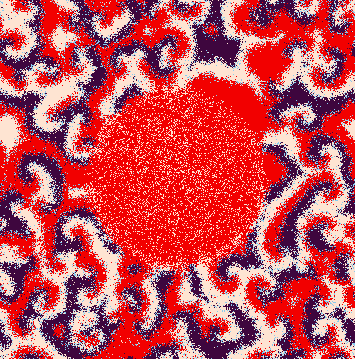}
        \caption{}\label{fig3d}
    \end{subfigure} \\
        \begin{subfigure}{.24\textwidth}
        \centering
        \includegraphics[width=33mm]{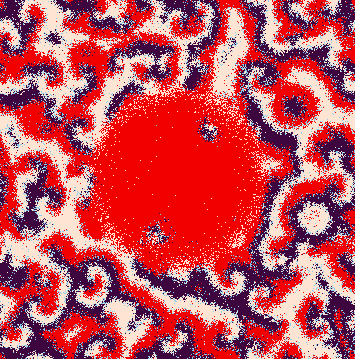}
        \caption{}\label{fig3e}
    \end{subfigure} %
       \begin{subfigure}{.24\textwidth}
        \centering
        \includegraphics[width=33mm]{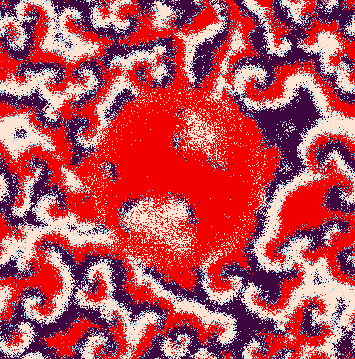}
        \caption{}\label{fig3f}
    \end{subfigure} %
   \begin{subfigure}{.24\textwidth}
        \centering
        \includegraphics[width=33mm]{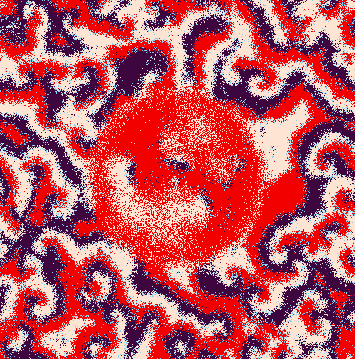}
        \caption{}\label{fig3g}
    \end{subfigure} 
   \begin{subfigure}{.24\textwidth}
        \centering
        \includegraphics[width=33mm]{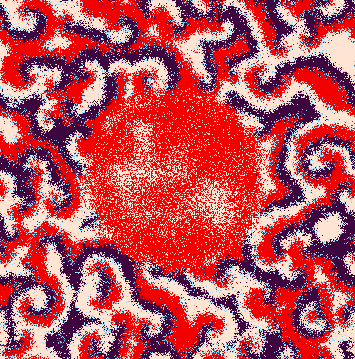}
        \caption{}\label{fig3h}
    \end{subfigure} 
\caption{Snapshots captured from a simulation of the rock-paper-scissors game with regional unevenness running in a lattice with $500^2$ grid points for a timespan of $5000$ generations.
Figures \ref{fig3a}, \ref{fig3b}, \ref{fig3c}, \ref{fig3d}, \ref{fig3e}, \ref{fig3f}, \ref{fig3g}, and \ref{fig3h} show the random initial conditions, and spatial organisation after $20$, $120$, $380$, $720$, $820$, $880$, and $1020$ generations, respectively. 
The colours follow the scheme in Fig. 1, with red, beige, and dark purple dots depicting individuals of species $1$, $2$, and $3$, respectively; empty spaces appear in blue dots. The simulation was implemented for $\mathcal{R}_{in}=125$, $\mathcal{R}_{out}=135$, $\chi=1.0$, and $s=r=m=1/3$. Video https://youtu.be/CgNH01Ajmj8 shows the transformation in the individuals' spatial organisation during the entire simulation.
}
  \label{fig3}
\end{figure*}
\begin{figure}
\centering
\includegraphics[width=89mm]{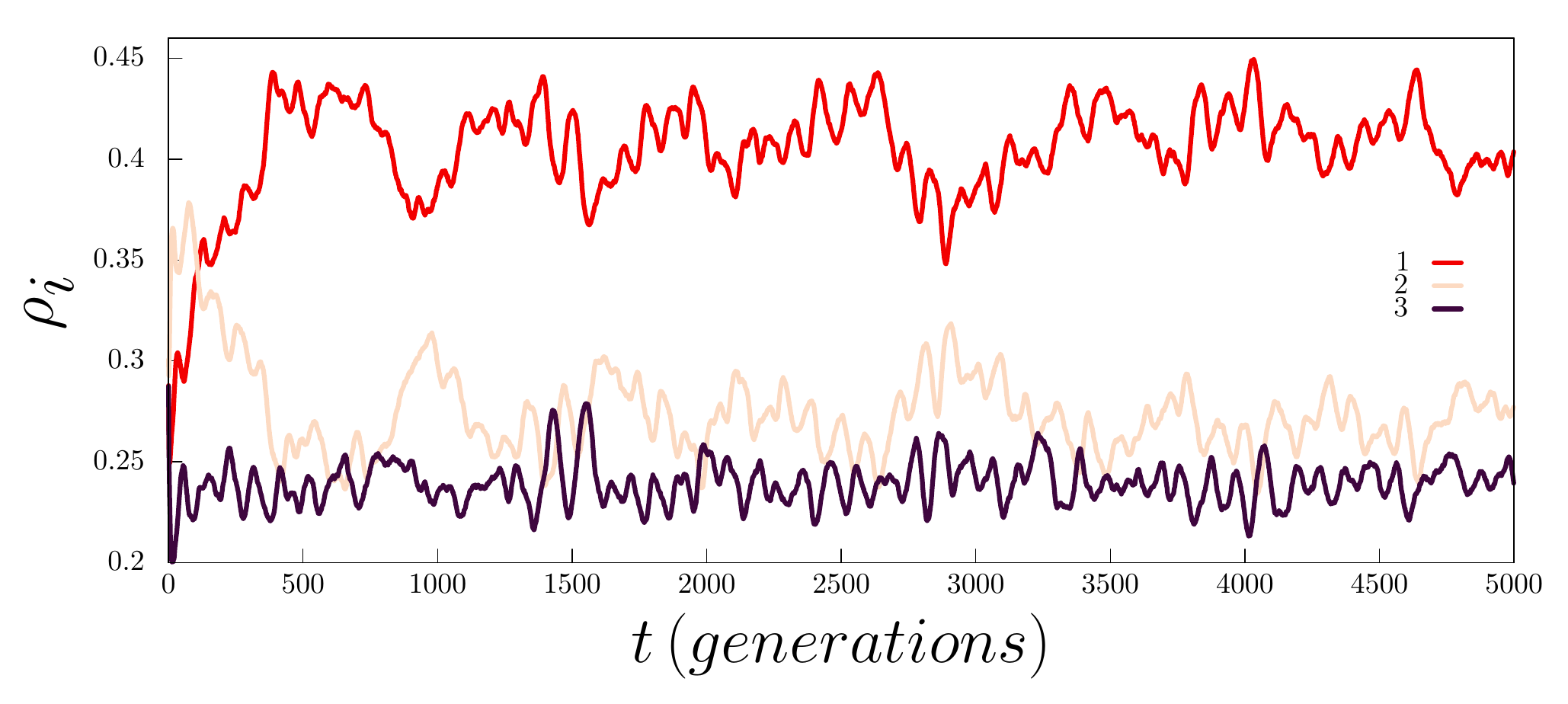}
\caption{Dynamics of the species densities during the course of the simulation in Fig.~\ref{fig3} and video https://youtu.be/5s5hl5WXuOI. The colours follow the scheme in Fig.~\ref{fig1}.}
	\label{fig4}
\end{figure}

\begin{figure}[t]
	\centering
	\includegraphics[width=89mm]{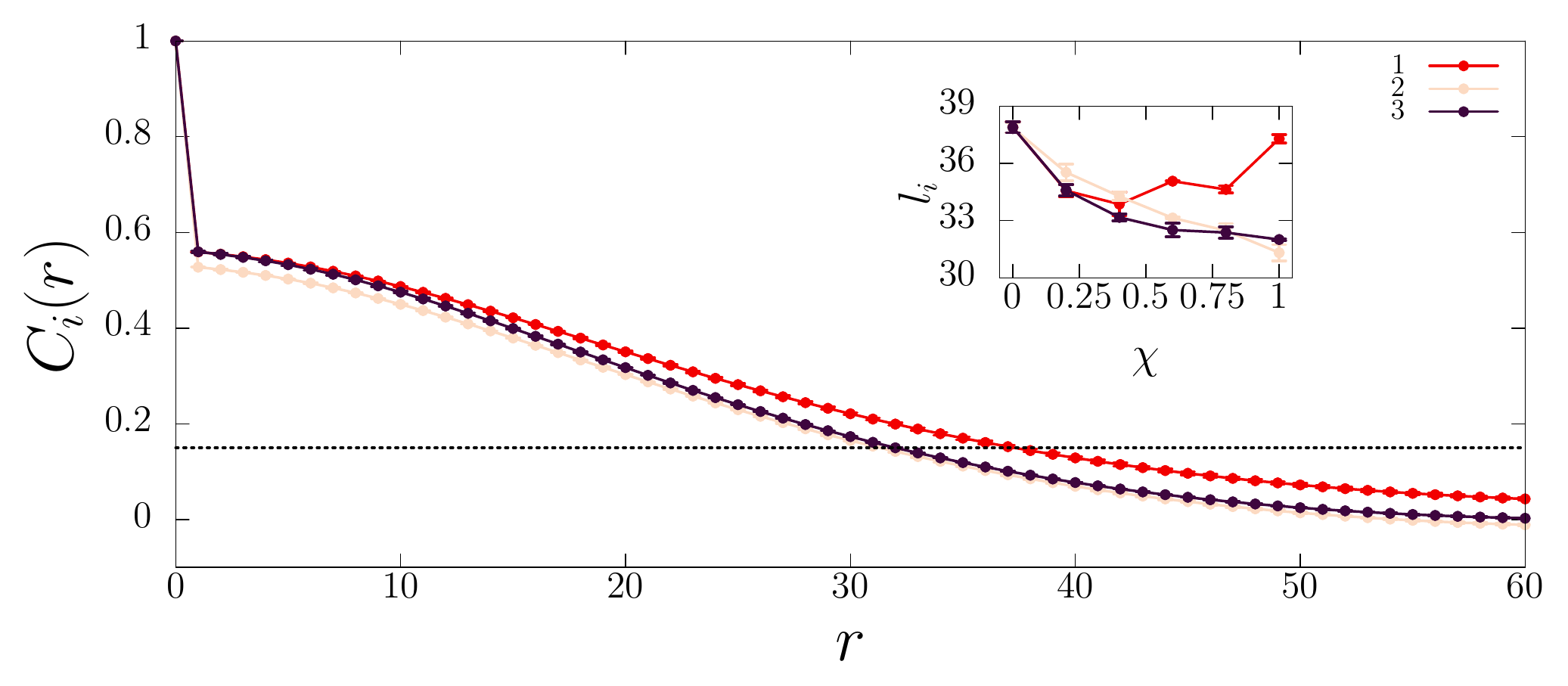}
    \caption{Autocorrelation functions $C_i$  as a function of the radial coordinate. The colours follow the scheme in Fig.~\ref{fig1}. The horizontal dashed black line indicates the threshold assumed to calculate the characteristic length depicted in the inset for a range of $\chi$. The error bars show the standard deviation by averaging the results obtained from sets of $100$ simulations with different random initial conditions.}
  \label{fig5-0}
\end{figure}

\section{Spatial patterns and population dynamics}
\label{sec3}

We first studied the pattern formation process starting from the random distribution of organisms depicted in Fig. \ref{fig3a}. The colours follow the scheme in Fig.~\ref{fig1}, where red, beige, and purple dots represent individuals of species $1$, $2$, and $3$, respectively; blue dots show the empty spaces. 
The realisation was performed in a square lattice with $500^2$ grid points; $\mathcal{S}_1$ is minimum within the central circular area of radius $\mathcal{R}_{in}=125$, while outside the disk of $\mathcal{R}_{out}=135$, the selection capacity of organisms of species $1$ is maximum. The interaction probabilities are $s=r=m=1/3$; the realisation ran for a timespan of $5000$ generations with $\chi=1.0$. The video https://youtu.be/CgNH01Ajmj8 shows the dynamics of spatial patterns during the entire simulation. 

Let us first focus on the pattern formation shown in the snapshots of Figs. \ref{fig3b}, \ref{fig3c}, and \ref{fig3d}, which were captured after $20$, $120$ and $380$ generations, respectively. The cyclic selection dominance inherent to the rock-paper-scissors game leads to the appearance of spiral waves for $\mathcal{R} \geq \mathcal{R}_{out}$, where organisms of every species have the same selection capacity.
As individuals of species $1$ suffer regional selection limitations, one has
the growth of regional population of species $2$ (beige) within the central region with radius $\mathcal{R} < \mathcal{R}_{out}$.
The consequence is the rapid local extinction of species $3$ (purple), as observed in the snapshot in Fig.~\ref{fig3b}. 
Subsequently, the number of individuals of species $1$ (red) rises due to the absence of organisms of species $3$. This leads to the gradual elimination of individuals of species $2$, despite the reduced selection capacity of
organisms of species $1$, as one sees in Fig.~\ref{fig3c} and \ref{fig3d}.
Therefore, species $1$, the weakest one, predominates in regions where environmental causes weaken it. \cite{weak-bacteria,weakest}.

Because the central circle is connected with the area without environmental selection limitations, the region is not permanently monopolised by species $1$. The outcomes show that, when the local density of organisms of species $2$ is low in the centre of the grid, the opposition to organisms of species $3$, which stochastically enter the region, is significantly reduced. 
This allows the creation of waves, as shown in Figs.~\ref{fig3e} to \ref{fig3h}, captured after $720$, $820$, $880$, and $1020$ generations, respectively.
The happens because: i) small groups of organisms of species $3$ (purple) conquer territories of species $1$ (red), creating invading waves in the central region; ii) following the cyclic game, organisms of species $2$ (beige) take the territory of that species $3$ had conquered, following them in the spreading waves; iii) the propagation of organisms of species $3$ is stochastically interrupted when they are caught by organisms of species $2$  chasing them; iv) the travelling waves are destroyed and individuals of species $1$ catch the remaining individuals of species $2$; v) the reoccupation of the central region by species $1$ holds until a new wave stochastically reaches the territory. 

Figure \ref{fig4} shows the dynamics of the species densities during the entire simulation shown in Fig.~\ref{fig3} - the colours follow the scheme in Fig.~\ref{fig1}.
As observed in the organisms' spatial configuration in Fig.~\ref{fig3}, as soon as the simulation commences, species $2$ gains transient preponderance. However, the prevalence in the spatial game is rapidly taken by species $1$, which prevails during the whole simulation. 
We also verified that, at certain points of the simulation, the density of individuals of species $3$ grows (for example, at $t \approx 2700$ generations), followed by an increase in $\rho_2$ and a decrease in $\rho_1$. This happens every time that, stochastically, the waves observed in Figs \ref{fig3e} invade the uneven selection central region.

\subsection{Spatial Autocorrelation Function}
We quantify the impact of the regional unevenness in the spatial correlation among individuals of the same species. For this purpose, we compute the spatial autocorrelation function $C_i(r)$, with $i=1,2,3$, in terms of radial coordinate $r$. We first introduce
the function $\phi_i(\vec{r})$ to describe the position $\vec{r}$ in the lattice occupied by individuals of species $i$. Then, we employ the mean value $\langle\phi_i\rangle$ to find the Fourier transform
\begin{equation}
\varphi_i(\vec{\kappa}) = \mathcal{F}\,\{\phi_i(\vec{r})-\langle\phi_i\rangle\}, 
\end{equation}
and the spectral densities
\begin{equation}
S_i(\vec{k}) = \sum_{k_x, k_y}\,\varphi_i(\vec{\kappa}).
\end{equation}

Using the normalised inverse Fourier transform, we compute the autocorrelation function for species $i$ as
\begin{equation}
C_i(\vec{r}') = \frac{\mathcal{F}^{-1}\{S_i(\vec{k})\}}{C(0)},
\end{equation}
which is written as a function of $r$ as
\begin{equation}
C_i(r') = \sum_{|\vec{r}'|=x+y} \frac{C_i(\vec{r}')}{min\left[2N-(x+y+1), (x+y+1)\right]}.
\end{equation}

We run a series of $100$ simulations with different initial conditions to calculate the mean autocorrelation function. 
The simulations were performed in lattices with $500^2$ grid sites, running until $t=5000$, using the same set of parameters used in the results shown in Figs.~\ref{fig3} and \ref{fig4}.
The red, beige, and dark purple lines in Fig.~\ref{fig5-0} show 
the outcomes for species 1, 2, and 3, respectively;
the error bars indicate the standard deviation. 

Assuming the threshold $C_i(l_i)=0.15$, we find the characteristic length scale of the spatial domains, $l_i$, with $i=1,2,3$, as the dashed black line represents it represents the threshold to calculate the characteristic scale $l_i$. 
The outcomes in the inset figure indicate that as $\chi$ grows, the unbalancing spatial autocorrelation of individuals of the same species accentuates. 
For $\chi=0$, which represents that all organisms of every species kill enemies with $50\%$ of their total capacity, the characteristic length scale of typical single-species domains (territories dominated by individuals of a single species) is approximately $l_1=l_2=l_3=38.5 \pm 0.023$. As $\chi$ increases, the gap between the average selection capacity between individuals of species $1$ and the others grows. For $\chi=1.0$, we discovered that results show that $l_1 \approx 1.19 l_2 \approx 1.16 l_3$.

\section{Effects of regional unevenness on species densities}
\label{sec4}


\begin{figure}
 \centering
       \begin{subfigure}{.47\textwidth}
        \centering
        \includegraphics[width=73mm]{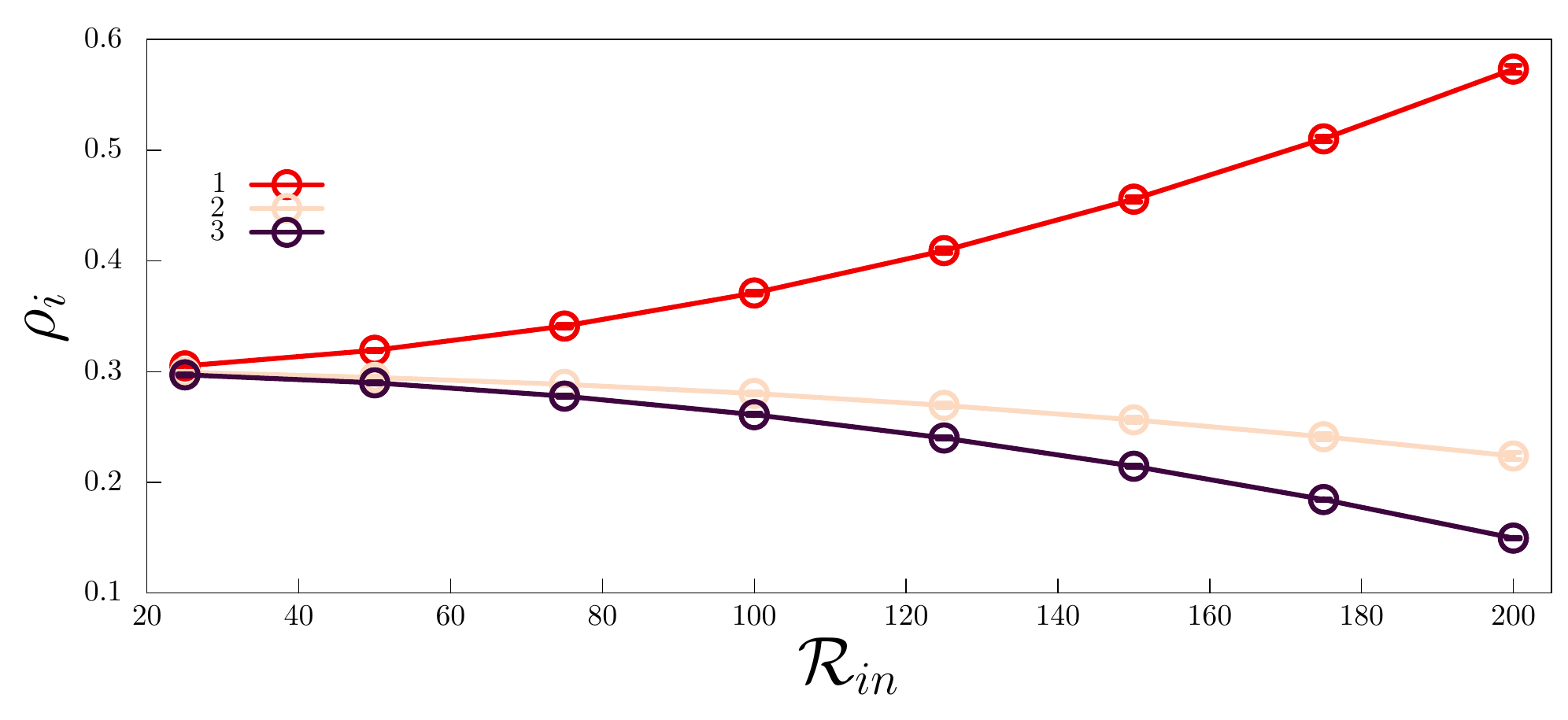}
        \caption{}\label{fig5a}
    \end{subfigure}
           \begin{subfigure}{.47\textwidth}
        \centering
        \includegraphics[width=73mm]{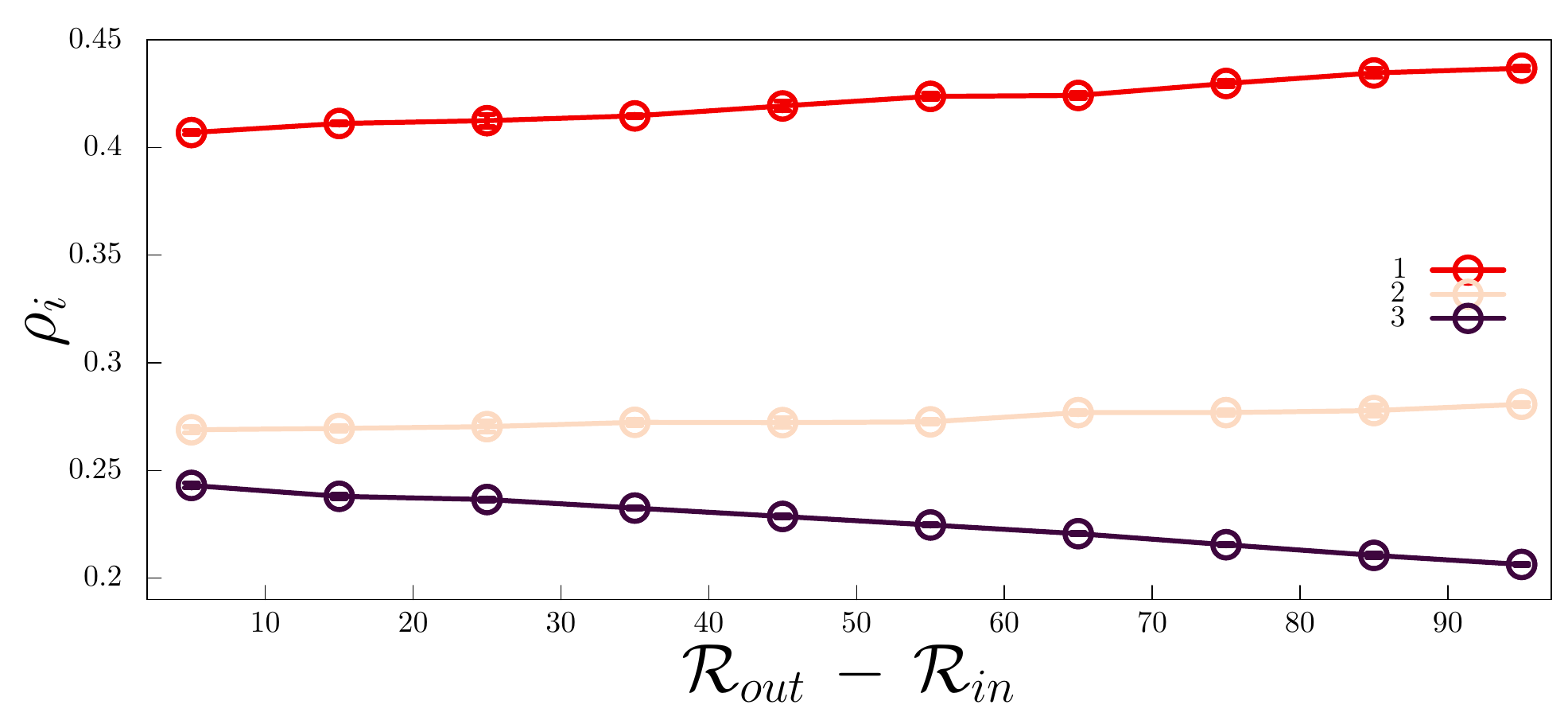}
        \caption{}\label{fig5b}
    \end{subfigure}\\
           \begin{subfigure}{.47\textwidth}
        \centering
        \includegraphics[width=73mm]{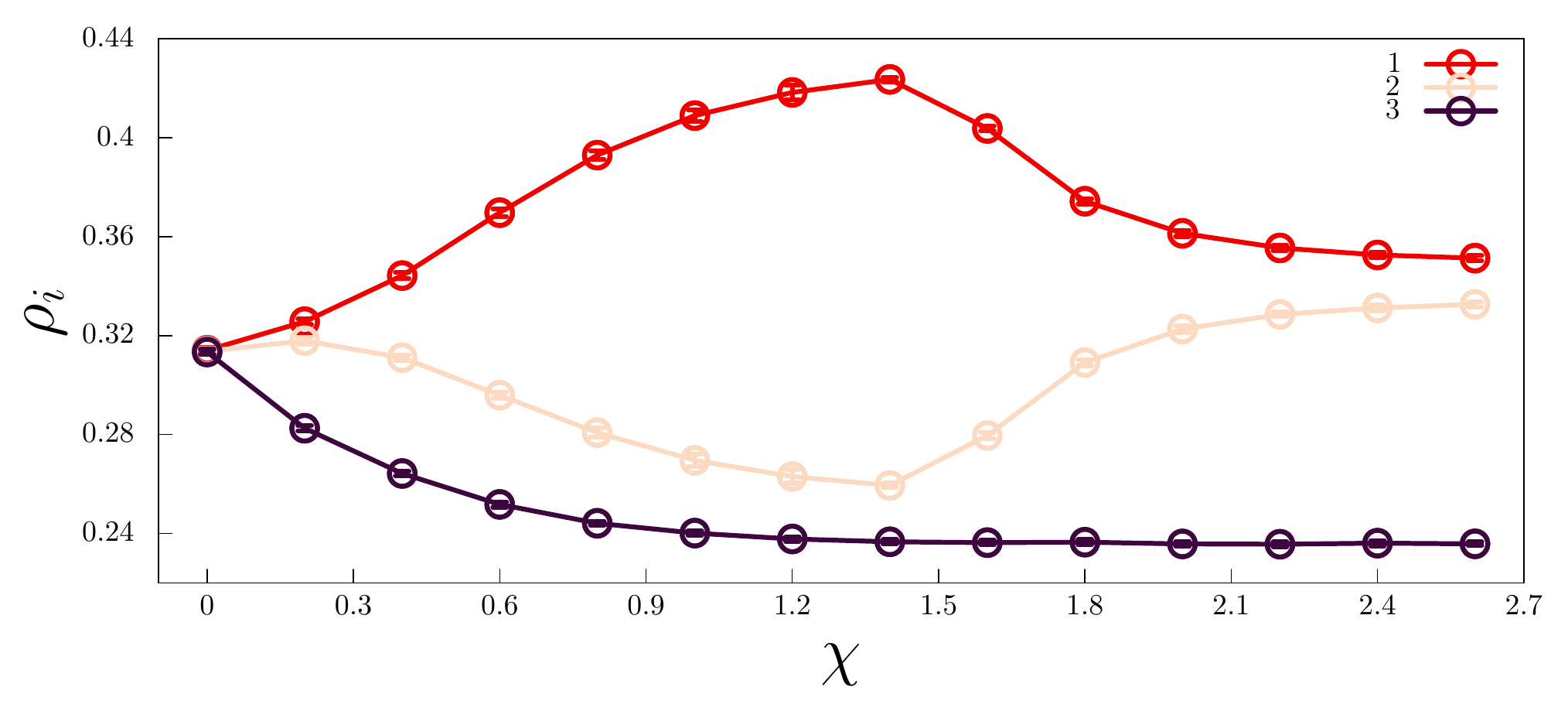}
        \caption{}\label{fig5c}
    \end{subfigure}
\caption{Species densities in the rock-paper-scissors model with regional unevenness. Figures \ref{fig5a} and \ref{fig5b} shows the dependence of $\rho_i$ on the radius of the selection-limiting radius $\mathcal{R}_{in}$ (for $\mathcal{R}_{out}-\mathcal{R}_{in}=10$ and $\chi=1.0$) and the width of the fringe transition area, $\mathcal{R}_{out}-\mathcal{R}_{in}$ (for $\mathcal{R}_{in}=125$ and $\chi=1.0$), respectively. Figure \ref{fig5c} depict the species densities for a range of transition parameter $\chi$ (for $\mathcal{R}_{in}=125$ and $\mathcal{R}_{out}-\mathcal{R}_{in}=10$). The results were obtained by averaging sets of $100$ simulations, running until $5000$ generations; the error bars show the standard deviation, and colours follow the scheme in Fig.\ref{fig1}.}
  \label{fig5}
\end{figure}

Now, we aim to quantify the influence of the geometric parameters
$\mathcal{R}_{out}$, $\mathcal{R}_{in}$, and the transition parameter $\chi$ in the species densities $\rho_i$.
For this purpose, we ran groups of $100$ simulations in lattices with $500^2$ grid points for a timespan of $5000$ generations; 
the interaction probabilities are $s=r=m=1/3$.
We compute the average $\rho_i$ considering the data from the second simulation half to avoid the initial fluctuation inherent to the initial pattern formation interval. The mean $\rho_i$ are shown in Fig.~\ref{fig5}; the colours follow the scheme in Fig.~\ref{fig1}, and the error bars indicate the standard deviation. 

The experiments are organised as follows: 
\begin{enumerate}
\item
We fixed $\mathcal{R}_{out}-\mathcal{R}_{in}=10$ and $\chi=1.0$, and performed sets of simulations for $25 \leq \mathcal{R}_{in} \leq 200$, with intervals of $\Delta \mathcal{R}_{in} = 25$; the results are depicted in Fig.~\ref{fig5a}.
\item
Assuming $\mathcal{R}_{in}=125$ and $\chi=1.0$, we ran simulations for $5 \leq (\mathcal{R}_{out} - \mathcal{R}_{in}) \leq 95$, with intervals of $\Delta (\mathcal{R}_{out} - \mathcal{R}_{in})= 10$; the outcomes appear in Fig.~\ref{fig5b}.
\item
For fixed $\mathcal{R}_{in}=125$ and $\mathcal{R}_{out}-\mathcal{R}_{in}=10$, we performed a series of realisations for 
$0 \leq \chi \leq 2.6$, with intervals of $\Delta \chi= 0.4$; the results are shown in Fig.~\ref{fig5c}.
\end{enumerate}

Our outcomes provide evidence that the species which is weakened by regional selection difficulties predominates irrespective of the topological characteristics of the selection-limiting region and transition parameter.
Figure \ref{fig5a} shows that the average fraction of the territory occupied by species $1$ increases as the area where organisms are affected grows, accentuating the reduction of $\rho_2$ and $\rho_3$.
Similarly, the positive effect in the population of species $1$ is observed as the transition fringe area grows, controlling a more significant fraction of the lattice for larger $\mathcal{R}_{out}-\mathcal{R}_{in}$, as shown in Fig. \ref{fig5b}.

Figure \ref{fig5c} shows an interesting effect introduced by the transition parameter in the species densities $\rho_1$. Accordingly, for $\chi=0$, there is no difference between the selection capacity of individuals of species $1$ and the others: $\mathcal{S}_1=\mathcal{S}_2=\mathcal{S}_3=1/2$; thus, every species occupies, on average, the same fraction of the territory. For $0 < \chi \leq 1.4$, $\rho_1$ grows while $\rho_2$ and $\rho_3$ drop. However, for $\chi>1.4$, the average fraction of the lattice dominated by individuals of species $1$ decreases, with the population of species $2$ rising; for $\chi \geq 2.0$, the variation in
$\rho_1$ and $\rho_3$ becomes smoother, with species $1$ prevailing in the spatial game.
\section{Coexistence Probability}
\label{sec5}
We investigate how the regional selection unevenness in the rock-paper-scissors game impacts species coexistence. For this purpose, we ran sets of $1000$ simulations in lattices with $100^2$ grid sites, for a timespan of $10000$ generations, considering $ 0.05\,<\,m\,<\,0.95$ in intervals of $ \Delta\, m\, =\,0.05$. For each simulation, selection and reproduction probabilities are given by $s\,=\,r\,=\,(1-m)/2$. 
We define coexistence as if at least one individual of every species is present at the end of the simulation, i.e., $I_i (t=10000) \neq 0$, for $i=1,2,3$.  Therefore, coexistence probability is the fraction of the simulations not resulting in extinction of any species.

\begin{figure}
 \centering
        \begin{subfigure}{.47\textwidth}
        \centering
        \includegraphics[width=73mm]{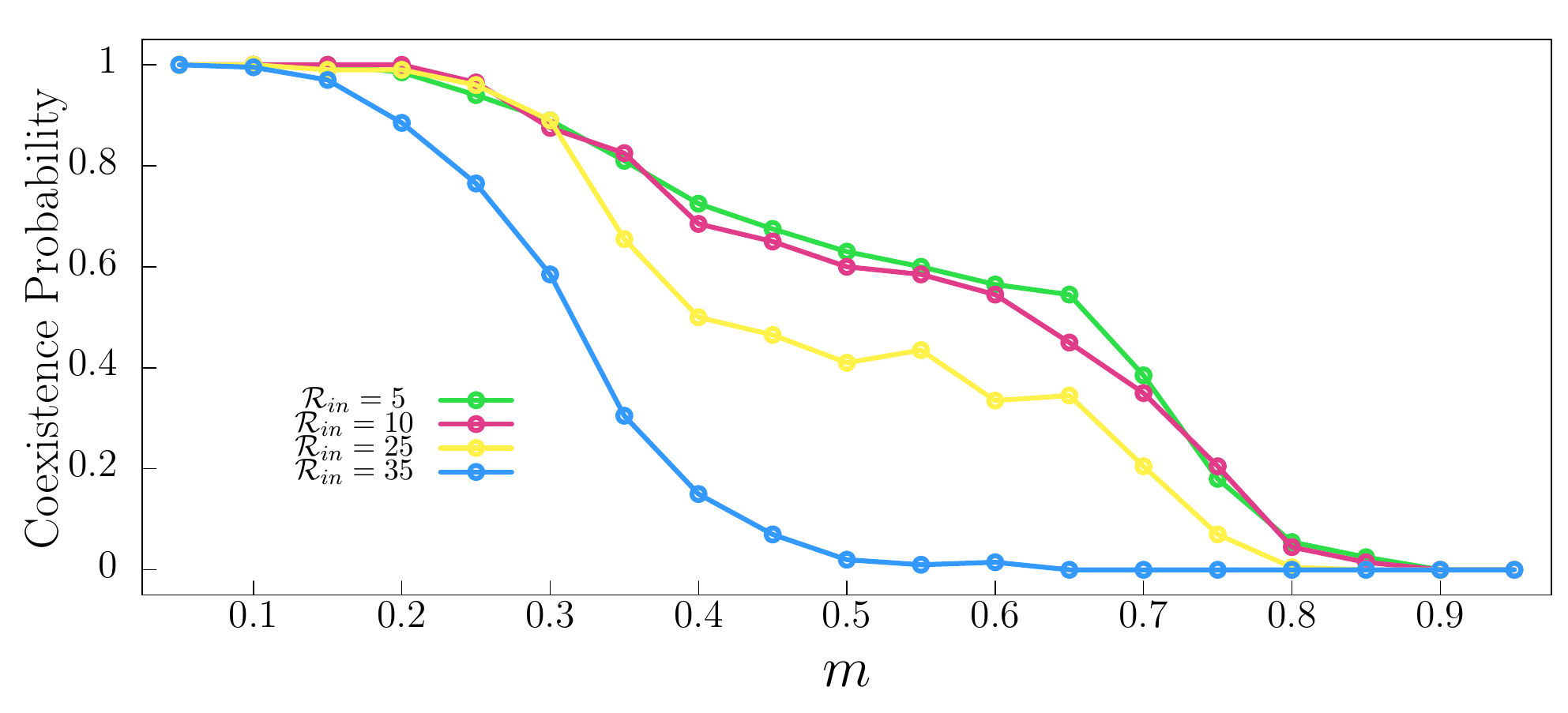}
        \caption{}\label{fig6a}
    \end{subfigure}
       \begin{subfigure}{.47\textwidth}
        \centering
        \includegraphics[width=73mm]{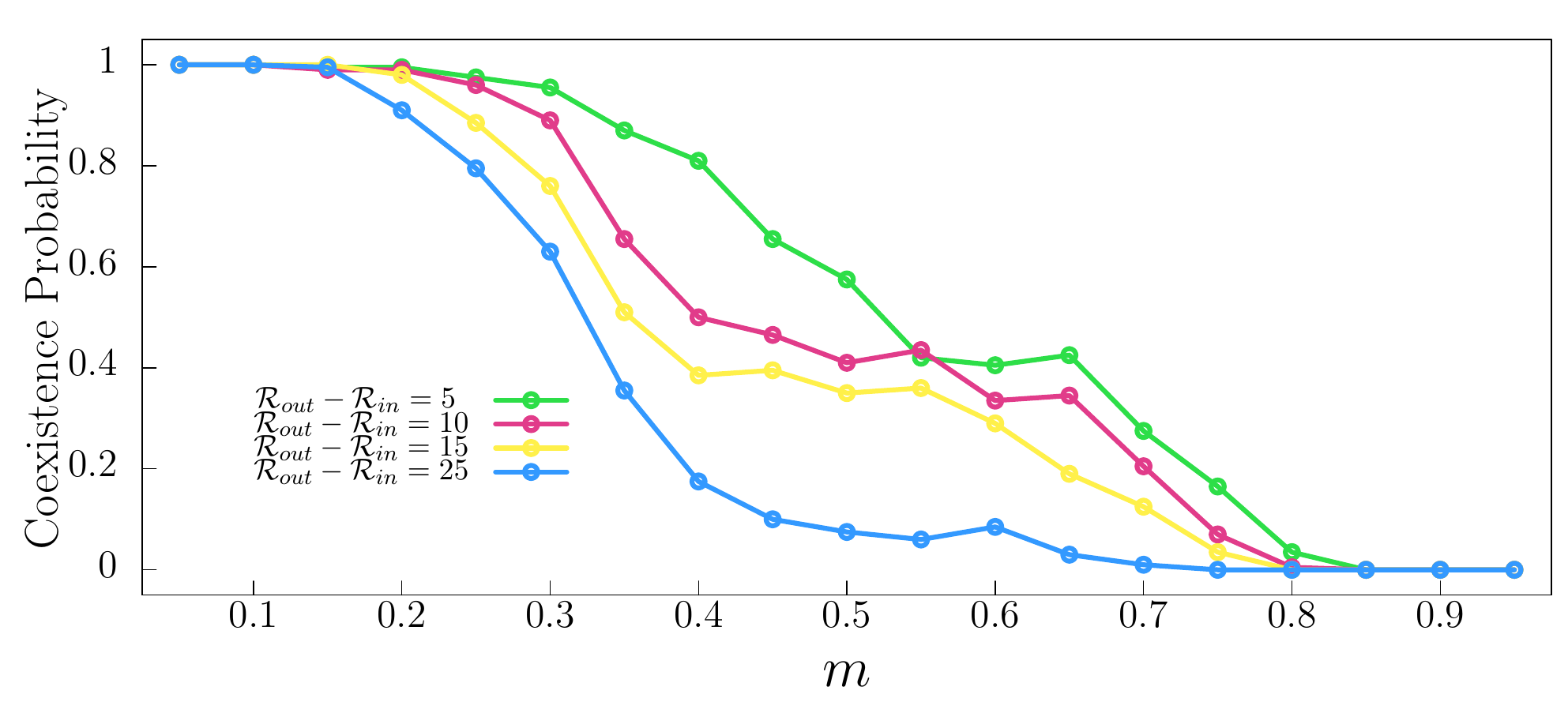}
        \caption{}\label{fig6b}
    \end{subfigure}\\
           \begin{subfigure}{.47\textwidth}
        \centering
        \includegraphics[width=73mm]{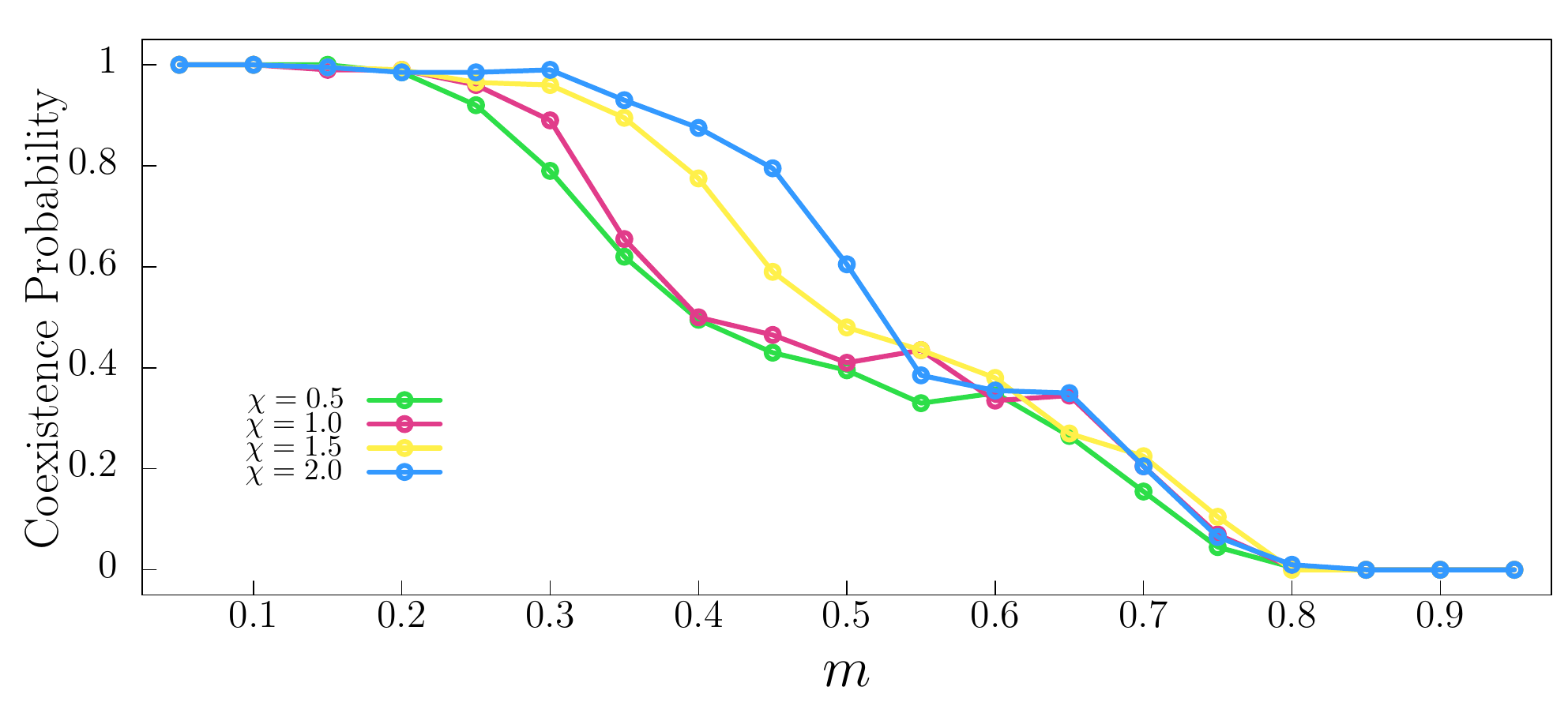}
        \caption{}\label{fig6c}
    \end{subfigure}
\caption{Coexistence probability as a function of the mobility probability $m$. Figure \ref{fig6a} shows the effects of radius of the central uneven region $\mathcal{R}_{in}$ on jeopardising biodiversity, for $\mathcal{R}_{out}-\mathcal{R}_{in}=10$ and $\chi=1.0$.
The influence of the fringe $\mathcal{R}_{out}-\mathcal{R}_{in}$ in coexistence probability is depicted in Fig.~\ref{fig6b}, for
$\mathcal{R}_{in}=25$ and $\chi=1.0$. Finally, the impact of transition on biodiversity is shown in Fig. \ref{fig6c}, for $\mathcal{R}_{in}=25$ and $\mathcal{R}_{out}-\mathcal{R}_{in}=10$. The results were obtained by running $1000$ simulations in lattices with $100^2$ grid points running until $100^2$ generations.}
  \label{fig6}
\end{figure}

Figure \ref{fig6a} shows how $\mathcal{R}_{in}$ affects biodiversity, for a fixed $\mathcal{R}_{in}-\mathcal{R}_{out}=10$ and $\chi=1.0$.
Green, red, yellow and blue lines depict the results for $\mathcal{R}_{in}=5$, $\mathcal{R}_{in}=10$, $\mathcal{R}_{in}=25$, and $\mathcal{R}_{in}=35$. Overall, the larger the radius of the selection limiting region, the more threatened the biodiversity. This happens because increasing the area where organisms of species $1$ are weakened means extending the territory where they predominate. Thus, the probability of species $2$ and $3$ disappearing rises as $\mathcal{R}_{in}$ grows. Furthermore,
the results show that the coexistence probability is approximately the same for short $\mathcal{R}_{in}$, as indicated by the green and red lines. However, biodiversity is significantly jeopardised for large $\mathcal{R}_{in}$. In the case of $\mathcal{R}_{in}=35$,
where the area of the central region represents $38\%$ of the lattice, biodiversity is lost even if organisms move with intermediate mobility probability; namely, coexistence probability  approaches zero for $m \geq 0.5$, as depicted by the blue line.

The outcomes in Figure \ref{fig6b} unveil the role of the transition fringe area. The experiment ran for fixed $\mathcal{R}_{in}=25$ and $\chi=1.0$; green, red, yellow, and blue lines show how coexistence is affected for $\mathcal{R}_{out}-\mathcal{R}_{in}=5$, $\mathcal{R}_{out}-\mathcal{R}_{in}=10$ , $\mathcal{R}_{out}-\mathcal{R}_{in}=15$, and $\mathcal{R}_{out}-\mathcal{R}_{in}=25$, respectively. Generally speaking, the larger the transition fringe area, the more extended the region where species $1$ is weakened, thus, accentuating the chances of biodiversity loss. 
For $\mathcal{R}_{out}-\mathcal{R}_{in}=25$, the chances of species coexisting decrease significantly for $m\geq 0.5$, as depicted by the blue line.

Finally, Figure \ref{fig6c} shows what happens if $\chi$ varies, for fixed $\mathcal{R}_{in}=25$ and $\mathcal{R}_{out}-\mathcal{R}_{in}=10$. The green, red, yellow, and blue lines show the outcomes for $\chi=0.5$, $\chi=1.0$, $\chi=1,5$, and $\chi=2.0$, respectively. 
We found that the more rapid the crossover transition between the minimum and maximum $\mathcal{S}_1$, the less threatened the biodiversity. This happens because although for small $\chi$, the average selection capacity of species $1$ is lower than the other species, the difference between them decreases. For example, in the central selection uneven region, for $\chi=1.0$, one has $\mathcal{S}_1 =0.268$, and $\mathcal{S}_2 =\mathcal{S}_3= 0.732$, while for $\chi=4.0$, $\mathcal{S}_1 =0.018$, and $\mathcal{S}_2 =\mathcal{S}_3= 0.982$. As for $\mathcal{R} \geq \mathcal{R}_{out}$, one has $\mathcal{S}_1 =\mathcal{S}_2=\mathcal{S}_3$, we conclude that for small $\chi$, the average difference in selection capacity decreases; thus, reducing the average unevenness selection in the cyclic competition. As the system is less unbalanced, the risk of biodiversity loss is dropped. However,
for low $\chi$, the average selection capacity is reduced for all species, so individuals' average reproduction and mobility rates rise. A larger effective mobility probability introduces an extra jeopardising effect on biodiversity, which is more jeopardised for small $\chi$. This effect becomes less significant as $\chi$ grows, with the selection unevenness in the central region being responsible for the reduction in the coexistence probability.

\section{Conclusions}
\label{sec6}

Our investigation implements a stochastic simulation of the spatial version of the rock-paper-scissors, where organisms of one out of the species face a regional reduction in their selection capacity.
The results of uneven rock-paper-scissors models in the literature refer to the case where one species is weaker because of an external factor, like a spreading disease that equally affects all organisms of one out of the species \cite{weakest,uneven,PedroWeak,Weak4,AVELINO2022111738}. Our study, in contrast, focuses on selection restrictions limited to a geographic area of the grid.

Running stochastic simulations, we found that the weakest species predominates in the regionally unbalanced cyclic game, independent of the size of the selection-limiting region and the intensity of the transition between the area within and outside the geographic area. 
Our outcomes show that in the case of species $i$ being the only surviving inside the selection-limiting region, transitory waves of organisms of species $i-1$ followed by individuals of species $i+1$ sporadically enter the region. The intruders initially proliferate but, affected by the unbalanced local cyclic selection game rules, are quickly destroyed \cite{MENEZES2022111903}. 

Our findings revealed that, as the size of the selection-limiting area grows, the average selection capacity of affected species drops. As this species becomes, on average, weaker, the predominance in the cyclic game accentuates, controlling a more significant fraction of the territory. Moreover, 
our results show that the advantage in the cyclic game is more significant if the transition between the limiting-selection area and the outer spatial positions is not sharp, but with selection capacity being reduced to zero in the inner grid points.

Regarding biodiversity, we quantified the effects of regional unevenness on the coexistence probability for organisms with a wide range of mobility. Overall, our outcomes show that the higher the mobility probability, the more jeopardised the biodiversity \cite{Reichenbach-N-448-1046}. More, the coexistence probability is more impacted as the selection-limiting region grows or the transition of selection capacity between the inner and outer grid is smoother. 

Our choice of species $1$ to be regionally affected is arbitrary because organisms of every species interact with the same probabilities (selection, reproduction, and mobility) in areas where environmental changes have not occurred - far from the lattice centre.
This means that spatial patterns and biodiversity are equally impacted if only species 2 or 3 regionally suffer a reduction in selection capacity. Furthermore, since the individuals are randomly distributed in the initial spatial configuration, our conclusions hold even for slightly unbalanced species densities. 
Our results may be useful to biologists to understand the biodiversity in regions where organisms' competition rules are affected by local environmental conditions.

\section*{Acknowledgments}
We thank CNPq, ECT, Fapern, and IBED for financial and technical support.

\section*{References}
\bibliographystyle{iopart-num}
\bibliography{ref}

\providecommand{\newblock}{}
\begin{thebibliography}{10}
\expandafter\ifx\csname url\endcsname\relax
  \def\url#1{{\tt #1}}\fi
\expandafter\ifx\csname urlprefix\endcsname\relax\def\urlprefix{URL }\fi
\providecommand{\eprint}[2][]{\url{#2}}

\bibitem{ecology}
Begon M, Townsend C~R and Harper J~L 2006 {\em Ecology: from individuals to
  ecosystems\/} (Oxford: Blackwell Publishing)

\bibitem{Nature-bio}
Purvis A and Hector A 2000 {\em Nature\/} {\bf 405} 212--2019

\bibitem{darwin}
Darwin C and Kebler L 1859 {\em On the origin of species by means of natural
  selection, or, The preservation of favoured races in the struggle for life\/}
  (London: J. Murray)

\bibitem{island1}
Losos J~B and Pringle R~M 2011 {\em Nature\/} {\bf 475}(7355) 1476--4687

\bibitem{bioislands}
MacArthur R~H and Wilson E~O 2001 {\em The Theory of Island Biogeography\/}
  (Princeton: Princeton University Press)

\bibitem{CHESSON1985263}
Chesson P~L 1985 {\em Theoretical Population Biology\/} {\bf 28} 263--287

\bibitem{doi:10.1086/323589}
Shurin J~B and Allen E~G 2001 {\em The American Naturalist\/} {\bf 158}
  624--637

\bibitem{https://doi.org/10.1890/02-0528}
Palmer T~M 2003 {\em Ecology\/} {\bf 84} 2843--2855

\bibitem{https://doi.org/10.1002/ece3.1980}
Yurkowski D~J, Ferguson S, Choy E~S, Loseto L~L, Brown T~M, Muir D~C~G,
  Semeniuk C~A~D and Fisk A~T 2016 {\em Ecology and Evolution\/} {\bf 6}
  1666--1678

\bibitem{10.1093/aobpla/plw081}
Crous C~J, Burgess T~I, Le~Roux J~J, Richardson D~M, Slippers B and Wingfield
  M~J 2016 {\em AoB PLANTS\/} {\bf 9}

\bibitem{geographic}
Wahle R, Brown C and Hovel K 2013 {\em Bulletin of Marine Science\/} {\bf 89}
  189--212

\bibitem{10.3389/fmars.2020.567758}
McMahan M~D, Sherwood G~D and Grabowski J~H 2020 {\em Frontiers in Marine
  Science\/} {\bf 7}

\bibitem{plantar}
Grevstad F~S and Klepetka B~W 1992 {\em Oecologia\/} {\bf 92} 399--404

\bibitem{urbanisation}
Concepción E~D, Moretti M, Altermatt F, Nobis M~P and Obrist M~K 2015 {\em
  Oikos\/} {\bf 124} 1571--1582

\bibitem{size}
da~Silva F~R, de~Moraes G~J, Lesna I, Sato Y, Vasquez C, Hanna R and Sabelis
  Maurice W~Janssen A 2016 {\em BioControl\/} {\bf 61} 681--689

\bibitem{climate}
Araújo M~B and Rahbek C 2006 {\em Science\/} {\bf 313} 1396--1397

\bibitem{climate2}
Williams J~J and Newbold T 2020 {\em Diversity and Distributions\/} {\bf 26}
  76--92

\bibitem{Coli}
Kerr B, Riley M~A, Feldman M~W and Bohannan B~J~M 2002 {\em Nature\/} {\bf 418}
  171

\bibitem{bacteria}
Kirkup B~C and Riley M~A 2004 {\em Nature\/} {\bf 428} 412--414

\bibitem{Allelopathy}
Durret R and Levin S 1997 {\em J. Theor. Biol.\/} {\bf 185} 165--171

\bibitem{Directional1}
Albertson R~C, Streelman, T J and Kocher T~D 2003 {\em Proc. Nat. Acad. Sci.\/}
  {\bf 100} 5252–5257

\bibitem{Directional2}
Quinn T~P, Hodgson S, Flynn L, Hilborn R and Rogers D~E 2007 {\em Ecol.
  Appl.\/} {\bf 17} 733–739

\bibitem{lizards}
Sinervo B and Lively C~M 1996 {\em Nature\/} {\bf 380} 240--243

\bibitem{Extra1}
Volkov I, Banavar J~R, Hubbell S~P and Maritan A 2007 {\em Nature\/} {\bf 450}
  45

\bibitem{Reichenbach-N-448-1046}
Reichenbach T, Mobilia M and Frey E 2007 {\em Nature\/} {\bf 448} 1046--1049

\bibitem{Szolnoki-JRSI-11-0735}
Szolnoki A, Mobilia M, Jiang L~L, Szczesny B, Rucklidge A~M and Perc M 2014
  {\em Journal of The Royal Society Interface\/} {\bf 11}

\bibitem{Moura}
Moura B and Menezes J 2021 {\em Scientific Reports\/} {\bf 11} 6413

\bibitem{Anti1}
Menezes J 2021 {\em Phys. Rev. E\/} {\bf 103}(5) 052216

\bibitem{anti2}
Menezes J and Moura B 2021 {\em Phys. Rev. E\/} {\bf 104}(5) 054201

\bibitem{MENEZES2022101606}
Menezes J, Rangel E and Moura B 2022 {\em Ecological Informatics\/} {\bf 69}
  101606

\bibitem{PhysRevE.97.032415}
Avelino P~P, Bazeia D, Losano L, Menezes J, de~Oliveira B~F and Santos M~A 2018
  {\em Phys. Rev. E\/} {\bf 97}(3) 032415

\bibitem{Avelino-PRE-86-036112}
Avelino P~P, Bazeia D, Losano L, Menezes J and Oliveira B~F 2012 {\em Phys.
  Rev. E\/} {\bf 86}(3) 036112

\bibitem{pairwise1}
Szolnoki A, Vukov J and Perc M~c~v 2014 {\em Phys. Rev. E\/} {\bf 89}(6) 062125

\bibitem{Nagatani2018}
Cheng H, Yao N, Huang Z~G, Park J, Do Y and Lai Y~C

\bibitem{ham}
Bazeia D, Menezes J, de~Oliveira B~F and Ramos J~G~G~S 2017 {\em Europhysics
  Letters\/} {\bf 119} 58003

\bibitem{PhysRevE.99.052310}
Avelino P~P, Menezes J, de~Oliveira B~F and Pereira T~A 2019 {\em Phys. Rev.
  E\/} {\bf 99}(5) 052310

\bibitem{TENORIO2022112430}
Tenorio M, Rangel E and Menezes J 2022 {\em Chaos, Solitons \& Fractals\/} {\bf
  162} 112430

\bibitem{tanimoto2}
Kabir K~A and Tanimoto J 2021 {\em Applied Mathematics and Computation\/} {\bf
  394} 125767

\bibitem{park22}
Mohd M~H and Park J 2021 {\em Chaos, Solitons \& Fractals\/} {\bf 153} 111579

\bibitem{SCHREIBER20131}
Schreiber S~J and Killingback T~P 2013 {\em Theoretical Population Biology\/}
  {\bf 86} 1--11

\bibitem{neigh}
Bazeia D, Bongestab M and {de Oliveira} B 2022 {\em Physica A: Statistical
  Mechanics and its Applications\/} {\bf 587} 126547

\bibitem{directionp}
Avelino P~P, de~Oliveira B~F and Silva J~V~O 2020 {\em Europhysics Letters\/}
  {\bf 132} 48003

\bibitem{weakest}
Frean M and Abraham E~R 2001 {\em Proc. R. Soc. Lond. B.\/} {\bf 268}
  1323--1327

\bibitem{uneven}
Menezes J, Moura B and Pereira T~A 2019 {\em Europhysics Letters\/} {\bf 126}
  18003

\bibitem{PedroWeak}
Avelino P~P, de~Oliveira B~F and Trintin R~S 2019 {\em Phys. Rev. E\/} {\bf
  100}(4) 042209

\bibitem{Weak4}
Avelino P~P, de~Oliveira B~F and Trintin R~S 2020 {\em Phys. Rev. E\/} {\bf
  101}(6) 062312

\bibitem{AVELINO2022111738}
Avelino P, {de Oliveira} B and Trintin R 2022 {\em Chaos, Solitons \&
  Fractals\/} {\bf 155} 111738

\bibitem{Park2017}
Park J, Do Y, Jang B and Lai Y~C 2017 {\em Scientific Reports\/} {\bf 7}
  2045--2322

\bibitem{PARKCHAOS}
Park J, Do Y and Jang B 2018 {\em Chaos\/} {\bf 28} 113110

\bibitem{PhysRevE.105.014215}
Islam S, Mondal A, Mobilia M, Bhattacharyya S and Hens C 2022 {\em Phys. Rev.
  E\/} {\bf 105}(1) 014215

\bibitem{MENEZES2022101028}
Menezes J, Rodrigues S and Batista S 2022 {\em Ecological Complexity\/} {\bf
  52} 101028

\bibitem{10.1063/5.0106165}
Menezes J, Batista S, Tenorio M, Triaca E and Moura B 2022 {\em Chaos: An
  Interdisciplinary Journal of Nonlinear Science\/} {\bf 32} 123142

\bibitem{MENEZES2023113290}
Menezes J and Barbalho R 2023 {\em Chaos, Solitons \& Fractals\/} {\bf 169}
  113290

\bibitem{Lotka}
Lotka A~J 1920 {\em Journal of the American Chemical Society\/} {\bf 42}
  1595--1599

\bibitem{Volterra}
Volterra V (1931) {\em Lecons dur la Theorie Mathematique de la Lutte pour la
  Vie\/} $1^\mathrm{st}$ ed (Gauthier-Villars, Paris)

\bibitem{leonard}
May R~M and Leonard W~J 1975 {\em SIAM J. Appl. Math.\/} {\bf 29} 243--253

\bibitem{LV-ML}
Avelino P~P, de~Oliveira B~F and Trintin R~S 2022 {\em Phys. Rev. E\/} {\bf
  105}(2) 024309

\bibitem{driven}
Bazeia D, Ferreira M~J~B and Oliveira B F~and S~A 2021 {\em Scientific
  Reports\/} {\bf 11}(1) 12512

\bibitem{weak-bacteria}
Liao M~J, Miano A and Nguyen C~B 2020 {\em Nature Communications\/} {\bf 11}
  2041--1723

\bibitem{MENEZES2022111903}
Menezes J and Moura B 2022 {\em Chaos, Solitons \& Fractals\/} {\bf 157} 111903

\end{thebibliography}
\end{document}